\documentclass[journal]{IEEEtran}
\usepackage{amsmath}
\usepackage{graphicx}
\usepackage{multirow}
\usepackage{nomencl}
\usepackage{cite}
\usepackage{graphicx,dblfloatfix}
\usepackage{balance} 
\usepackage{flushend}
\usepackage{epstopdf} 
\usepackage{algorithm}
\usepackage{algorithmic}
\usepackage{setspace}
\usepackage{color}
\IEEEoverridecommandlockouts

\usepackage{eso-pic}
\newcommand\AtPageUpperMyright[1]{\AtPageUpperLeft{%
		\put(\LenToUnit{0cm},\LenToUnit{-1cm}){%
			\parbox{0.68\textwidth}{\raggedleft\fontsize{9}{11}\selectfont #1}}%
}}%
\newcommand{\conf}[1]{%
	\AddToShipoutPictureBG*{%
		\AtPageUpperMyright{#1}
	}
}    
\conf{Conference on Innovative Smart Grid Technology (ISGT), Washington, DC, 2018.}
\begin{document}
	\renewcommand*\footnoterule{}
	\title{Model Predictive BESS Control for Demand Charge Management and PV-Utilization Improvement} 
	\author{M. Ehsan Raoufat,~\textit{Student Member,~IEEE},
		Babak Asghari,~\textit{Member,~IEEE,}
		Ratnesh Sharma,~\textit{Member,~IEEE} \vspace{-0.35cm}
		\thanks{M. Ehsan Raoufat is with the Min H. Kao Department of Electrical Engineering and Computer Science, The University of Tennessee, Knoxville, TN 37996 USA (e-mail: mraoufat@utk.edu).}
		\thanks{Babak Asghari and Ratnesh Sharma are with the Energy Management Department, NEC Laboratories America Inc., Cupertino, CA, 95014 USA. }
	}
	\maketitle 
\begin{abstract}
\boldmath
Adoption of battery energy storage systems for behind-the-meters application offers valuable benefits for demand charge management as well as increasing PV-utilization. The key point is that while the benefit/cost ratio for a single application may not be favorable for economic benefits of storage systems, stacked services can provide multiple revenue streams for the same investment. Under this framework, we propose a model predictive controller to reduce demand charge cost and enhance PV-utilization level simultaneously. Different load patterns have been considered in this study and results are compared to the conventional rule-based controller.  The results verified that the proposed controller provides satisfactory performance by improving the PV-utilization rate between $60\%$ to $80\%$ without significant changes in demand charge (DC) saving. Furthermore, our results suggest that batteries can be used for stacking multiple services to improve their benefits. Quantitative analysis for PV-utilization as a function of battery size and prediction time window has also been carried out.
\end{abstract}
\renewcommand\IEEEkeywordsname{Index Terms}
\section*{Nomenclature}
\onehalfspacing
\addcontentsline{toc}{section}{Nomenclature}
\begin{IEEEdescription}[\IEEEusemathlabelsep\IEEEsetlabelwidth{$X_{1}X_{2}X_{3}X_{4}X_5$}]
	\item[$\eta_{pv}$]         				 				{PV-utilization rate;}
	\item[$P^{pur}_{g}, P^{sell}_{g}$]       				{purchased/sold power from/to  grid, kW;}
	\item[$P_{load}, P_{pv}$]     	   	     				{Load/PV power, kW;}
	\item[$P_{b}^{cha}$, $P_{b}^{dis}$]      				{BESS charge/discharge power, kW;}
	\item[$P_{b}^{\max}$]      								{BESS charge/discharge power limit, kW;}
	\item[$SOC$]     	   	   				 				{BESS state-of-charge;}
	\item[$SOC_{\min}$,$SOC_{\max}$]     	   	   			{BESS state-of-charge limits;}
	\item[$E^{sell}_{_{BESS}}$, $E^{sell}_{_{No BESS}}$]    {Excess energy sold to the grid with or without BESS, kWh;}
	\item[$DCT$]     	   	   				 				{Demand charge threshold, kW;}
	\item[$\lambda_{DC}$]     	   	   				 		{Demand charge rate, \$/kW}
	\item[$C_{tp}$]     	   	   				 		    {Battery throughput cost, \$/kW}
\end{IEEEdescription} 
\singlespacing
\section{Introduction}
Introducing battery energy storage system (BESS) for behind-the-meter (BTM) application offers valuable benefits for both the utility and the customers \cite{ESS_values}. However, the popularity of these storage systems mainly depends on the extent to which they can provide valuable services at a reasonable cost. In most recent applications, BESS are typically assigned and dispatched according to a single primary objective such as demand charge (DC) management \cite{Z_Wang, D_Wu}, however, the benefits of the BESS to the utility and to the end users can be far greater. BTM energy storage systems for commercial \& industrial (C\&I) segments can also be available to generate additional revenues by delivering services such as frequency regulation \cite{Y_Shi}, ramp rate control \cite{B_P_Bhattarai} and PV-utilization \cite{E_Vrettos}. The key point is that while the benefit/cost ratio for a single application may not be favorable, stacked services can provide multiple revenue streams for the same investment.\\ \indent 
Recently, growing penetration of rooftop PV generation is causing operational challenges for utilities to keep the voltage within an acceptable range. Consequently, new studies have been focused on using BTM BESS to increase the local PV-utilization. From the utility's point of view, maximizing the local PV-utilization reduces voltage fluctuations, improves power quality, and minimizes the loss in distribution networks \cite{EuropeanCom, G_Litjens,M_Motalleb}. Additionally, C\&I customers with storage systems can benefit from incentive support programs for self-consumptions, prevent curtailment loss due to feed-in limitations, and avoid the reduced PV feed-in tariff while providing other services simultaneously. For instance, during a daily operation, a BTM  BESS can discharge to reduce the peak demand and also charge by absorb the excess PV generation locally. \\ \indent
The challenge in combining DC management and PV-utilization service lies in their opposing nature: DC reduction works better when BESS is fully charged and ready to shave the peak, while PV-utilization prefers a battery which is not fully charged and ready to store the excess generation. The work in \cite{B_P_Bhattarai} studied the influence of load patterns on PV-self consumption but did not provide any type of controllers. In \cite{C_J_C_Williams}, a ``delayed charging'' algorithm is proposed to increase the state of charge (SOC) every 15-minutes in a linear fashion. However, in the case of peak shaving event in early hours, there is not enough energy stored in the battery. Results of another study \cite{E_Vrettos} suggest using batteries and flexible loads to increase the PV utilization and several rule-based controllers have been also proposed. However, the problem of DC management has not been considered and the assumption of having controllable loads are often not true for C\&I customers.\\ \indent
In this paper, we propose a new model predictive controller (MPC) to reduce the peak power injection from the grid and increase the PV-utilization level for BTM applications simultaneously. These services can be provided during the day through the optimal shift of load and PV generation in time. In our approach, peak shaving is of key concern and the MPC controller calculates the optimal charging/discharging guidelines for 15-minutes intervals which will then be sent to the real-time controller to ensure robust real-time operation. If it's necessary, charging and discharging profiles are overridden using a set of rules. In addition, the proposed MPC allows us to account for battery degradation and operational boundaries. The simulations are carried out using real data for different types of loads, battery sizes, and optimization horizons. The results show that the proposed method has satisfactory performance compared to conventional rule-based controllers. Moreover, our approach reduces the average annual SOC and has the added benefits of increasing the battery lifetime. \\ \indent
This paper is organized as follows. System description and modeling are demonstrated in Section II followed by a cooperative energy management solution to increase PV-utilization in Section III. Section IV presents the evaluation results using real measured data. Finally, concluding remarks are presented in Section V.
\section{System Description and Modeling}
\begin{figure}[!t]
	\centering
	\includegraphics[width=3.5in]{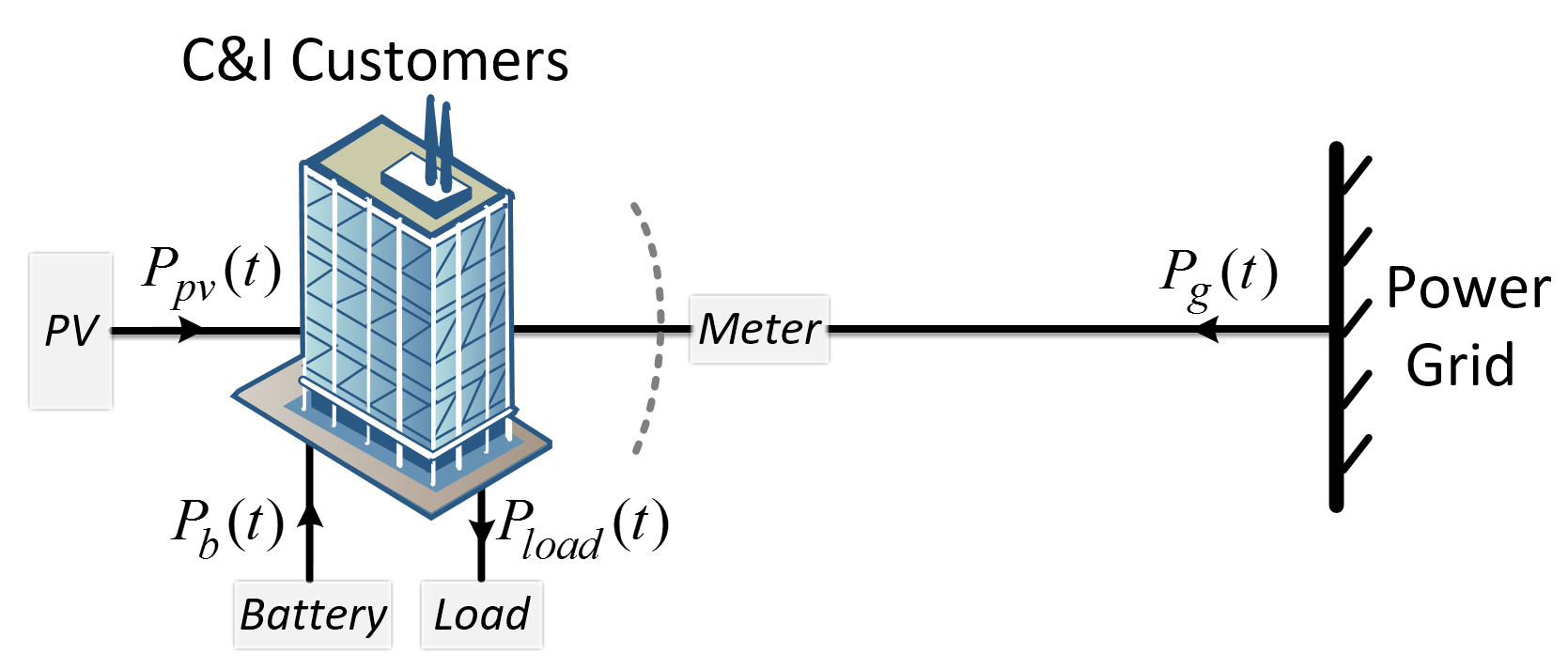}   \vspace{-0.4cm}
	\caption{System configuration for BTM applications.}
	\label{fig:BTM_Diagram}
\end{figure}

Schematic diagram of the system under consideration is demonstrated in Fig. \ref{fig:BTM_Diagram}. This system consists of a commercial or industrial load served by rooftop PV system complementing the main power grid and a battery energy storage system. The power balance equations are given as follows
\begin{IEEEeqnarray}{c}	
	P_{pv}(t) + P_{b}(t) + P_{g}(t) - P_{load}(t) = 0 \label{PBE}\\
	P_{g}(t) = P_{g}^{pur}(t) - P_{g}^{sell}(t) \\
	P_{b}(t) = P_{b}^{dis}(t) - P_{b}^{cha}(t)   \\
	0 \leq P_{b}^{dis}(t), P_{b}^{cha}(t) \leq P_{b}^{\max}
\end{IEEEeqnarray}	
In addition, the battery SOC evolves as follows
\begin{IEEEeqnarray}{c}
	SOC(t+1)=SOC(t) + P_{b}^{cha}(t) - P_{b}^{dis}(t) \\
	SOC_{\min} \leq SOC (t) \leq SOC_{\max}
\end{IEEEeqnarray}
\noindent The notations are as defined in the Nomenclature. The only control variable in (\ref{PBE}) is $P_b$ where others are disturbances as we have only considered uncontrollable loads. Note that the degradation of the battery that occurs during usage has not been taken into account in the above model.
\subsection{Parameters}
Two major parameters for BESS performance are investigated in this research; PV-utilization rate and DC cost. The PV-utilization rate $(\eta_{pv})$ is defined as one minus the ratio between the excess PV energy sold to the grid with and without BESS operation (which is measured in months or years).
\begin{IEEEeqnarray}{c}	
	\eta_{pv}= 1-\frac{ E^{sell}_{_{BESS}} }{ E^{sell}_{_{NoBESS}} }
\end{IEEEeqnarray}
The second parameter is DC cost $(J_{DC})$ in an electric bill which is calculated for an entire month by summing three components of anytime, partial peak, and peak power measured by the utility grid as follows
\begin{IEEEeqnarray}{c}	
	J_{DC}= \lambda_{DC}^{any} \times \underset{\forall t}{\max}[P_g(t)] + \lambda_{DC}^{partial} \times \underset{t \in p. p. t.}{\max}[P_g(t)] \nonumber \\
	\quad + \lambda_{DC}^{peak} \times \underset{t \in p. t.}{\max}[P_g(t)]
\end{IEEEeqnarray}
To improve the DC saving using conventional methods, the energy management system has been designed as a two layer architecture including monthly and daily layers as follows.
\subsection{Monthly layer, DC threshold generation}
The objective of this layer is to calculate the optimal peak grid powers to follow during the next billing cycle \cite{D_Wu}. These target grid powers are calculated based on 15-minutes interval and are called demand charge thresholds (DCTs). It should be noted that PV-utilization is a daily problem. However, demand charge is a monthly problem and has already been defined using historical data for the current month which is not the focus of this paper. As a result, during daily operation, the aggregated system needs to follow predefined thresholds calculated in the monthly layer. 
\subsection{Daily layer, conventional rule-based controller}
The daily layer is responsible for the continuous adjustment of battery and grid power based on real-time data. The rule-based control algorithm \ref{Alg: 1} is typically used in the daily layer to track DCT values generated in the monthly layer. However, this controller always charges the battery as long as demand is below the DCT and discharge the battery when demand is above the DCT. Consequently, the battery is fully charged for most of the time and it does not have any available capacity to capture the excess PV generation. In the next section, a new solution is proposed for the daily layer controller.
\begin{algorithm}[!t]
	\footnotesize
	\caption{Conventional rule-based controller:}
	Case 1: \quad if $(P_g(t) \leq DCT(t))$ \& $(SOC(t) \leq SOC_{\max})$
	\begin{itemize}
		\item[] \qquad \quad  $ P_{b}^{cha}(t) = DCT(t)-P_g(t) $
		\item[] \qquad \quad  $ P_{b}^{dis}(t) = 0 $
	\end{itemize}
	Case 2: \quad if $(P_g(t) > DCT(t))$ \& $(SOC(t) > SOC_{\min})$
	\begin{itemize}
		\item[] \qquad \quad $ P_{b}^{cha}(t) = 0 $
		\item[] \qquad \quad $ P_{b}^{dis}(t) = P_g(t)-DCT(t) $
	\end{itemize}
	Otherwise:
	\begin{itemize}
		\item[] \qquad \quad $ P_{b}^{cha}(t) = 0 $
		\item[] \qquad \quad $ P_{b}^{dis}(t) = 0 $
	\end{itemize}
	* Note that  $0 \leq P_{b}^{dis}(t), P_{b}^{cha}(t) \leq P_{b}^{\max}$.
	\label{Alg: 1}
\end{algorithm}
\section{Coordinated Energy Management Solution}
This section provides basic definitions and detailed optimization setup to address the demand charge management and PV-utilization problem simultaneously. 
\subsection{Model predictive controller}
We consider a finite time partitioned into $T$ discrete time intervals, indexed by $t \in \{1,2, \dots ,T \}$. A finite horizon constrained optimization problem is proposed as follows to find the optimal charging/discharging guidelines.
\begin{subequations}
	\label{eq:Opt1}
	\begin{align}
		& \underset{}{\text{min}}
		& & J=\sum_{t=1}^{T} P_{g}^{sell}(t) + C_{tp} (P_b^{cha}(t)+P_b^{dis}(t))\\ 
		& \text{s.t.}
		& & SOC_{req} \leq SOC (t) \leq SOC_{\max} \\
		& & & SOC(t+1)=SOC(t) + P_{b}^{cha}(t) - P_{b}^{dis}(t)  \\
		& & & P_{g}^{sell}(t)=P_{g}^{pur}(t) + P_{b}^{dis}(t) - P_{b}^{cha}(t) \nonumber  \\ & & &  \qquad \qquad + P_{pv}(t) - P_{load}(t) \\
		& & & P_g^{pur}(t) \leq DCT(t) \\
		& & & 0 \leq P_{b}^{dis}(t), P_{b}^{cha}(t) \leq P_{b}^{\max}
	\end{align}
\end{subequations}
\indent The objective function (\ref{eq:Opt1}a) minimizes the total excess PV generation sold to the grid and the battery degradation cost. Constraints (\ref{eq:Opt1}b) and (\ref{eq:Opt1}c) represent the battery SOC limits and dynamic equation, respectively. These constraints can be used to keep the SOC within the predefined boundaries in order to keep the energy storage healthier (longer lifetime) and operate within higher efficiency region. Constraint (\ref{eq:Opt1}d) is intended to take account the physics of the system, (\ref{eq:Opt1}e) guarantees that the DC thresholds will not be violated and (\ref{eq:Opt1}f) represents the battery operational boundaries. \\ \indent
The proposed method calculates optimal charging/discharging guidelines marked with a ``*'' as $P_{b}^{cha^*}(t)$ and $P_{b}^{dis^*}(t)$, respectively. This method consequently promotes the battery to discharge in hours before high PV generations or charging in hours before peak demands. However, the above optimization problem does not consider the case where DCT might be violated or SOC is lower than $SOC_{req}$ which may lead to infeasibility. Therefore, the overall optimization is reformulated using soft constraints to avoid these problems. \vspace{-0.4cm}
\begin{subequations}
	\label{eq:Opt2}
	\begin{align}
	& \underset{}{\text{min}}
	& & J=\sum_{t=1}^{T} P_{g}^{sell}(t) + C_{tp} \big( P_b^{cha}(t)+P_b^{dis}(t) \big) \nonumber \\ & & &  \qquad \qquad+ \alpha  \times SOC_{req}^{slack} + \beta \times DCT^{slack} \\ 
	& \text{s.t.}
	& & SOC_{req}-SOC_{req}^{slack} \leq SOC (t) \leq SOC_{\max} \\
	& & & SOC(t+1)=SOC(t) - P_{b}^{dis}(t) + P_{b}^{cha}(t) \\
	& & & P_{g}^{sell}(t)=P_{g}^{pur}(t) + P_{b}^{dis}(t) - P_{b}^{cha}(t)  \nonumber \\ & & &  \qquad \qquad + P_{pv}(t) - P_{load}(t) \\
	& & & P_g^{pur}(t) -DCT^{slack} \leq DCT(t) \\
	& & & 0 \leq P_{b}^{dis}(t), P_{b}^{cha}(t) \leq P_{b}^{\max}
	\end{align}
\end{subequations}
Variables $\alpha$ and $\beta$ are weightings for $SOC^{slack}_{req}$ and $DCT^{slack}$, respectively. The slack variables are added to avoid the hard constraint of threshold violations. This approach will provide guidelines to reduce the demand charge cost and increase the PV-utilization rate. Variables $P_{load}$ and $P_{pv}$ come from the estimated values of load and PV profiles for the next $T$ steps. The only variables that need to be chosen is $SOC_{req}$ which will be chosen based on the average minimum required battery capacity for peak shaving of the previous days.
\begin{algorithm}[t]
	\footnotesize
	\caption{Real-time controller:}
	Case 1: \quad if $(P_g(t) > DCT(t))$ \& $(SOC(t) > SOC_{\min})$
	\begin{itemize}
		\item[] \qquad \quad $ P_{b}^{cha}(t) = 0 $
		\item[] \qquad \quad $ P_{b}^{dis}(t) = P_g(t)-DCT(t) $
	\end{itemize}
	Case 2: \quad if $(P_g(t) < 0)$ \& $(SOC(t) < SOC_{\max})$
	\begin{itemize}
		\item[] \qquad \quad $ P_{b}^{cha}(t) = P_g(t) $
		\item[] \qquad \quad $ P_{b}^{dis}(t) = 0 $
	\end{itemize}
	Otherwise:
	\begin{itemize}
		\item[] \qquad \quad  $ P_{b}^{cha}(t) = P_{b}^{cha^*}(t) $
		\item[] \qquad \quad  $ P_{b}^{dis}(t) = P_{b}^{dis^*}(t) $
	\end{itemize}
	* Note that  $0 \leq P_{b}^{dis}(t), P_{b}^{cha}(t) \leq P_{b}^{\max}$.
	\label{Alg: 2}
\end{algorithm}
\subsection{Real-time Controller}
\begin{figure}[!t]  \vspace{-0.3cm}
	\centering
	\includegraphics[width=3in]{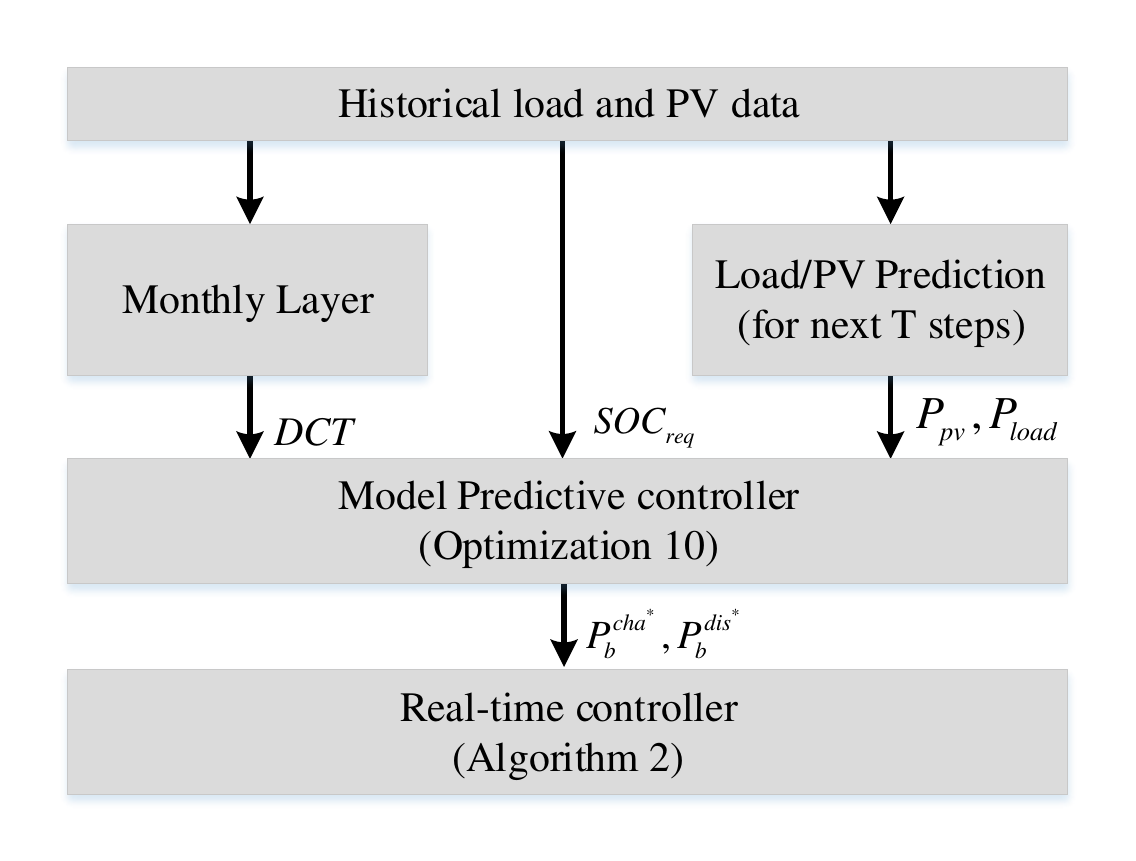}  \vspace{-0.4cm}
	\caption{Block diagram of the proposed MPC-based controller.}   \vspace{-0.3cm}
	\label{fig:Diagram}
\end{figure}

Here we propose a simple real-time control algorithm for the previous optimization problem. The previous controller calculates the optimal charging/discharging profiles for 15-minutes interval which will then be sent to the real-time controller Algorithm \ref{Alg: 2}. Nevertheless, even though those profiles are optimally calculated, they need additional adjustment in actual operation to overcome the forecasting errors and successful implementation in faster time scales (e.g. order of seconds). In this algorithm, discharging profile is overridden to avoid DCT violations in case the SOC is higher than the minimum value. Similarly, charging profile is overridden to increase the local PV-utilization where the SOC is lower than the predefined maximum value. Finally, the overall control algorithm is summarized in Fig. \ref{fig:Diagram}.
\begin{table}[!t]
	\caption{Demand Charge Rates from PG\&E.}
	\scriptsize
	\centering
	\renewcommand{\arraystretch}{1.2}
	\label{table:DCrate}
	\begin{tabular}{|c|c|c|c|}
		\hline
		\!\!\!\!	Demand Charge (DC) \!\!\!\!      & \!\! Rate (\$/kW) \!\! &  \!\!\!\! $1^{st}$ Time Window  \!\!\!\!  & \!\!\!\! $2^{nd}$ Time Window  \!\!\!\! \\ \hline  \hline
		Anytime (May-Oct.)                       & 17.44        &  0:00-24:00 & 0:00-0:00  \\ \hline
		\!\!\!\!	Partial peak (May-Oct.) \!\!\!\! & 0.50         &  8:30-12:00 & 18:00-21:30  \\ \hline
		Peak (May-Oct.)  	                     & 1.45         & 12:00-18:00 & 0:00-0:00  \\ \hline \hline
		Anytime  (Nov.-Apr.)                     & 17.44        &  0:00-24:00 & 0:00-0:00  \\ \hline
		\!\!\!\!	Partial peak (Nov.-Apr.)\!\!\!\! & 0.01         &  8:30-21:30 & 0:00-0:00  \\ \hline  
		Peak (Nov.-Apr.)	                     & 0.00         &  0:00-0:00  & 0:00-0:00  \\ \hline
	\end{tabular} 
\end{table}
\begin{table*}[!t]  \vspace{-0.3cm}
	\centering
	\caption{Yearly Simulation Results.}
	\label{TABLE:Year}
	\scriptsize
	\renewcommand{\arraystretch}{1.2}
	\begin{tabular}{|c|c|c|c|c|c|c|c|c|c|}
		\hline
		\multirow{2}{*}{Load Name}       & \multicolumn{1}{c|}{\multirow{2}{*}{No BESS}} & \multicolumn{8}{c|}{BESS}   \\ \cline{3-10} 
		& \multicolumn{1}{c|}{}          & \multicolumn{4}{c|}{Rule-based Controller}                              & \multicolumn{4}{c|}{MPC-based Controller}                                       \\ \cline{2-10} 
		& DC Cost (\$)   & DC Cost (\$)  & DC Saving ($\%$) & $SOC_{avg}$ & PV-util. ($\%$) & DC Cost (\$) & DC Saving ($\%$) & $SOC_{avg}$  &   PV-util. ($\%$)       \\ \hline
		Grocery         &   80330        &   65648       & 18.28           & 94.11      &  0.008      &  66337       & 17.42     & 57.20      &   71.15          \\ \hline
		Hospital        &   82660        &   72855       & 11.86           & 96.2       &  0          &  74906       &  9.38     & 63.86      &   82.38           \\ \hline
		Theater         &   75767        &   65035       & 14.17           & 97.7       &  0          &  65135       & 14.03     & 72.87      &   61.84           \\ \hline
	\end{tabular}
\end{table*}
\begin{figure*}[!t]
	\centering  \vspace{-0.15cm}
	\includegraphics[width=6.9in]{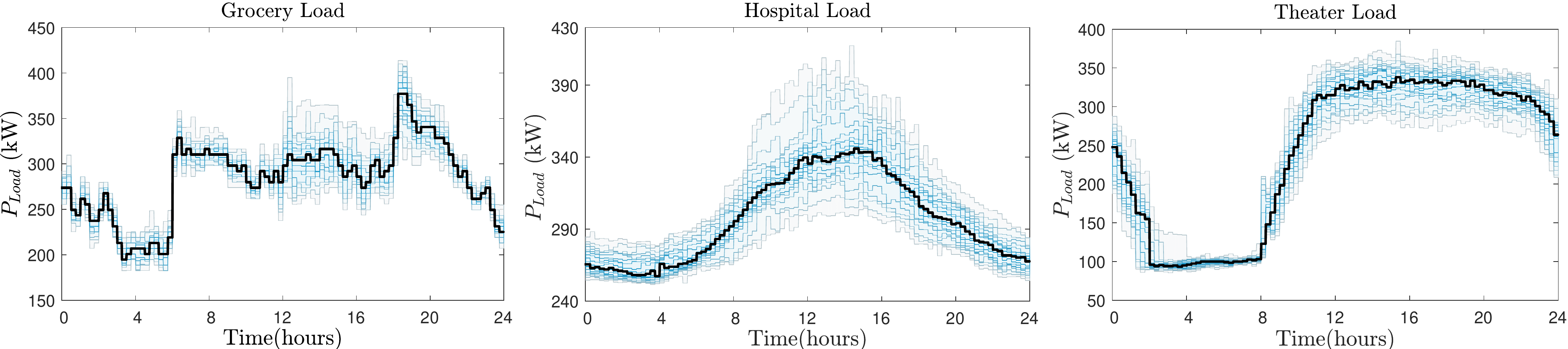}  \vspace{-0.25cm}
	\caption{Grocery, hospital and theater load profiles for each day in July, note that the grocery load profile has the highest variations.} 
	\label{fig:Load}
\end{figure*}
\begin{figure}[!t] 
	\centering   \vspace{-0.25cm}
	\includegraphics[width=3.351in]{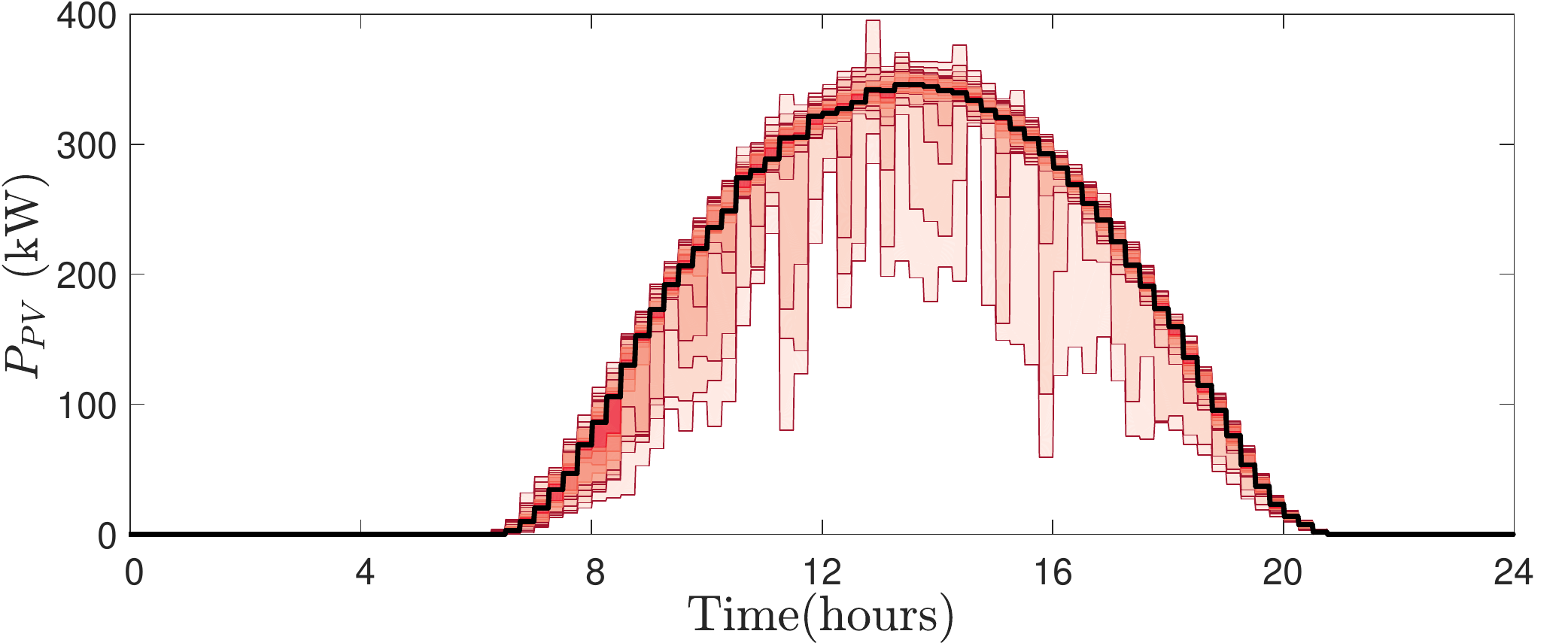}  \vspace{-0.2cm} 
	\caption{PV profile for each day in July.}   \vspace{-0.2cm}
	\label{fig:NECPV}
\end{figure}
\section{Simulation Results}
Extensive simulations are performed in C++ and performance of the proposed algorithm is demonstrated using real data for different load profiles, battery sizes, and optimization time horizons. The optimization problem (\ref{eq:Opt2}) is solved using GLPK software based on 15 minutes interval. However, it is worth mentioning that this algorithm can suit other time scales. In order to compare differences in electricity demand patterns, the peak loads were scaled to be $420$ kW and the PV penetration is assumed to be $90\%$ in all the cases. In this study, PV penetration is defined as the ratio of peak PV power to peak load power. The optimization time horizon is assumed to be 4-hours ($T=16$ steps) which can be changed based on the PV/load forecast accuracy and the battery size is chosen as $340$kWh/$710$kW, unless otherwise stated. The weighting functions and gains are chosen as follows: $\alpha=10$, $\beta=100$, and $C_{tp}=0.05$. Demand charge cost is calculated using rates from PG\&E markets for customers with renewable as presented in Table \ref{table:DCrate}.
\subsection{Case studies}
\begin{figure}[!t]
	\centering \vspace{-0.2cm}
	\includegraphics[width=2.9in]{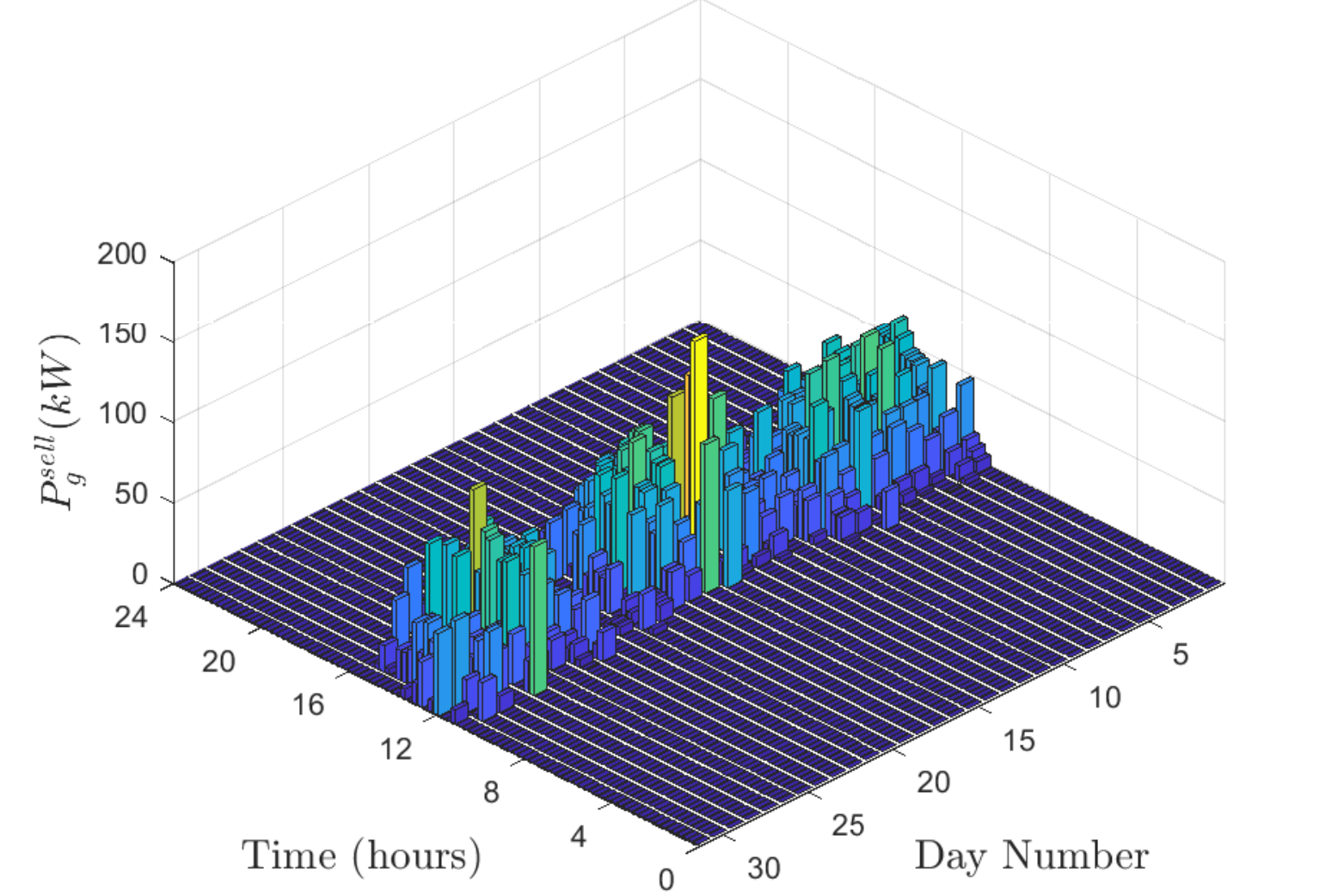} 
	\caption{Excess PV injection in July for grocery load while using the rule-based controller.}  \vspace{+0.3cm}
	\label{fig:Psell_RB}
	\centering \vspace{-0.2cm}
	\includegraphics[width=2.9in]{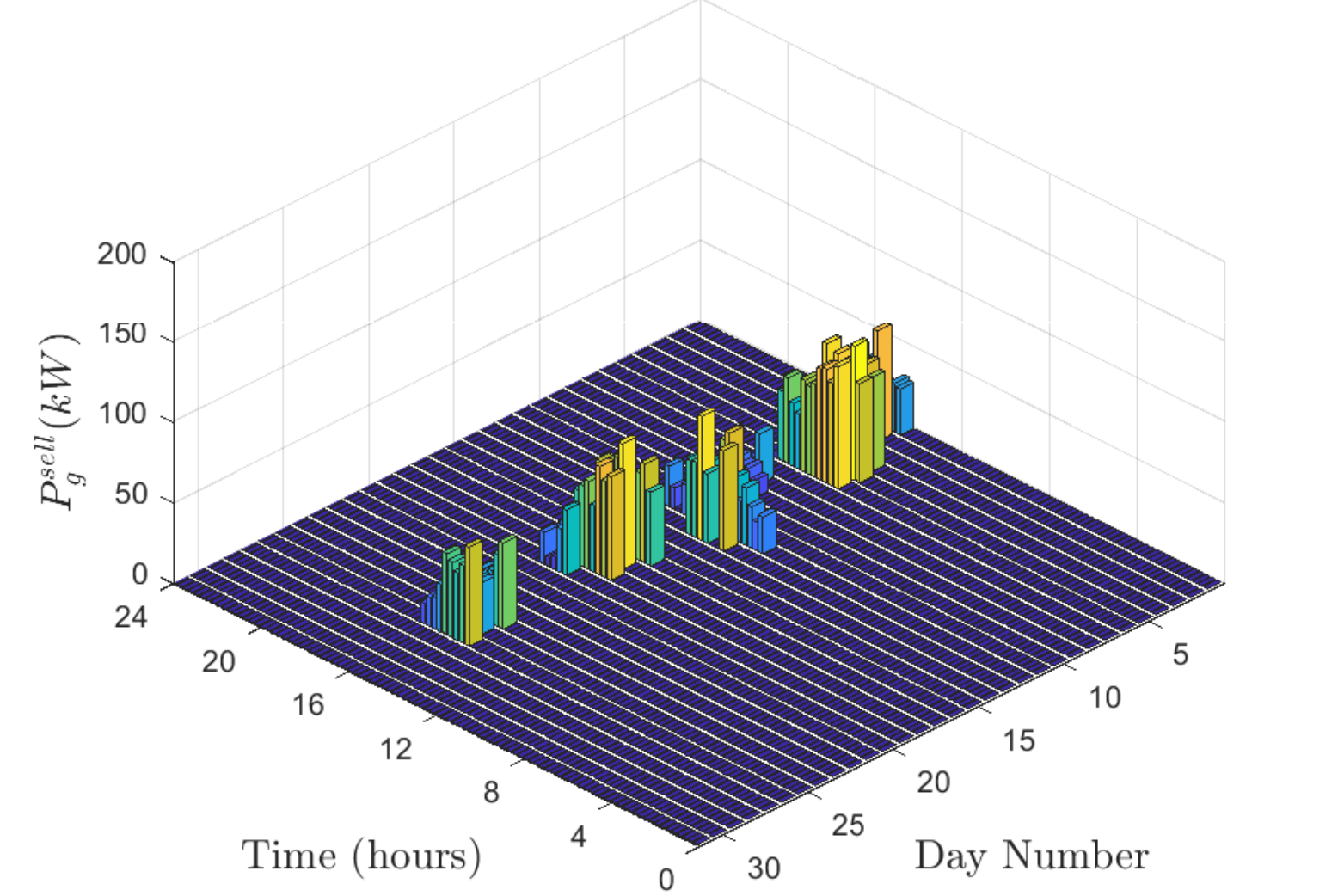} 
	\caption{Excess PV injection in July for grocery load while using the proposed MPC-based controller.} 
	\label{fig:Psell_OP}  \vspace{-0.2cm}
\end{figure}

Three separate case studies including Load profiles of a grocery store, a theater, and a hospital were considered to evaluate the proposed method. Load profiles for the month of July are shown in Fig. \ref{fig:Load}. It can be seen that grocery load profile is flatter in compare to theater profile with longer peak duration (happens late at nights) and hospital demand profile which is more aligned with PV generation pattern. The PV profile used in this study was measured at a fixed rooftop PV installation, profiles are illustrated in Fig. \ref{fig:NECPV} for the month of July with highest PV generation.\\ \indent 
\begin{figure}[!t]
	\centering \vspace{-0.5cm}
	\includegraphics[width=3.4in]{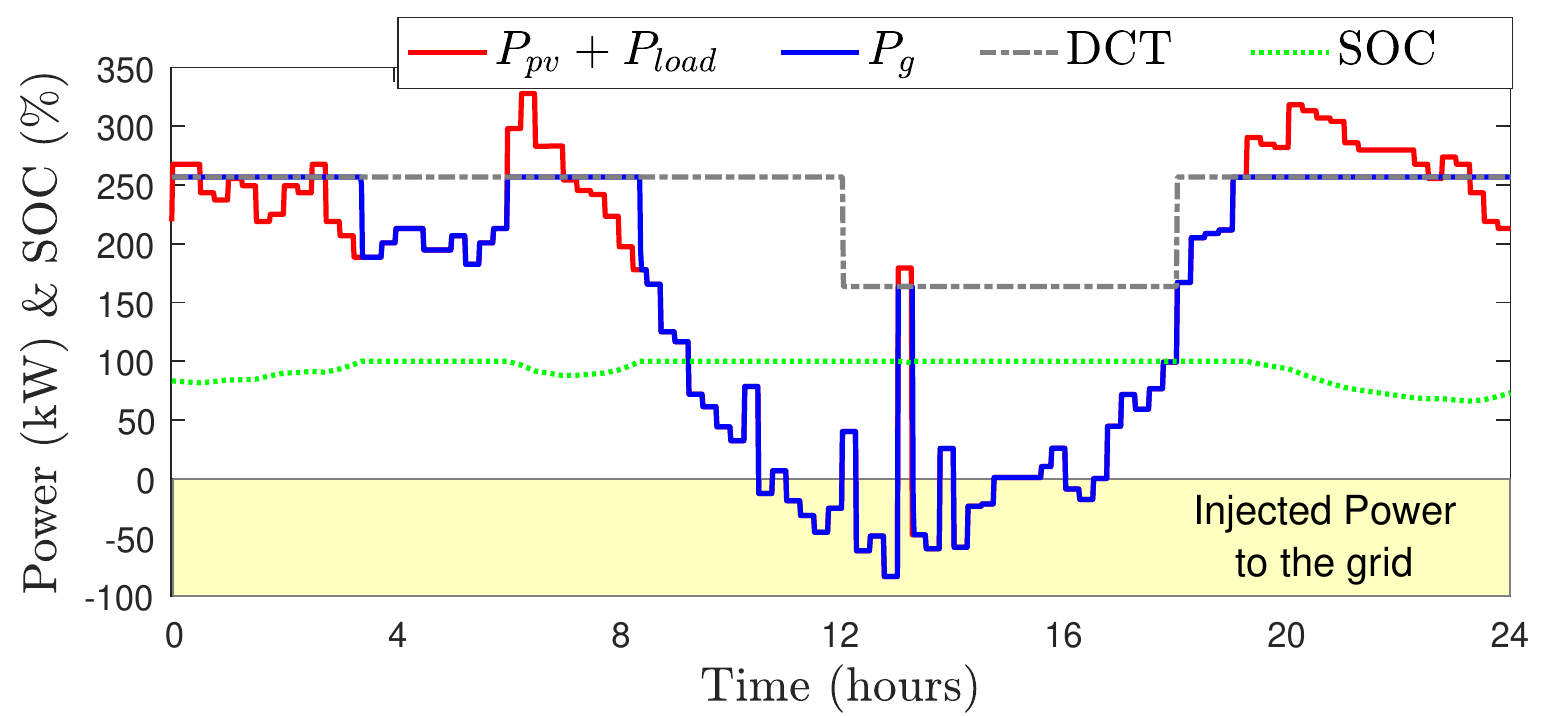}   \vspace{-0.2cm}
	\caption{Sample of a day in July for grocery load while using the rule-based controller.} 
	\label{fig:Daily_RB}
	\centering   \vspace{+0.2cm}
	\includegraphics[width=3.4in]{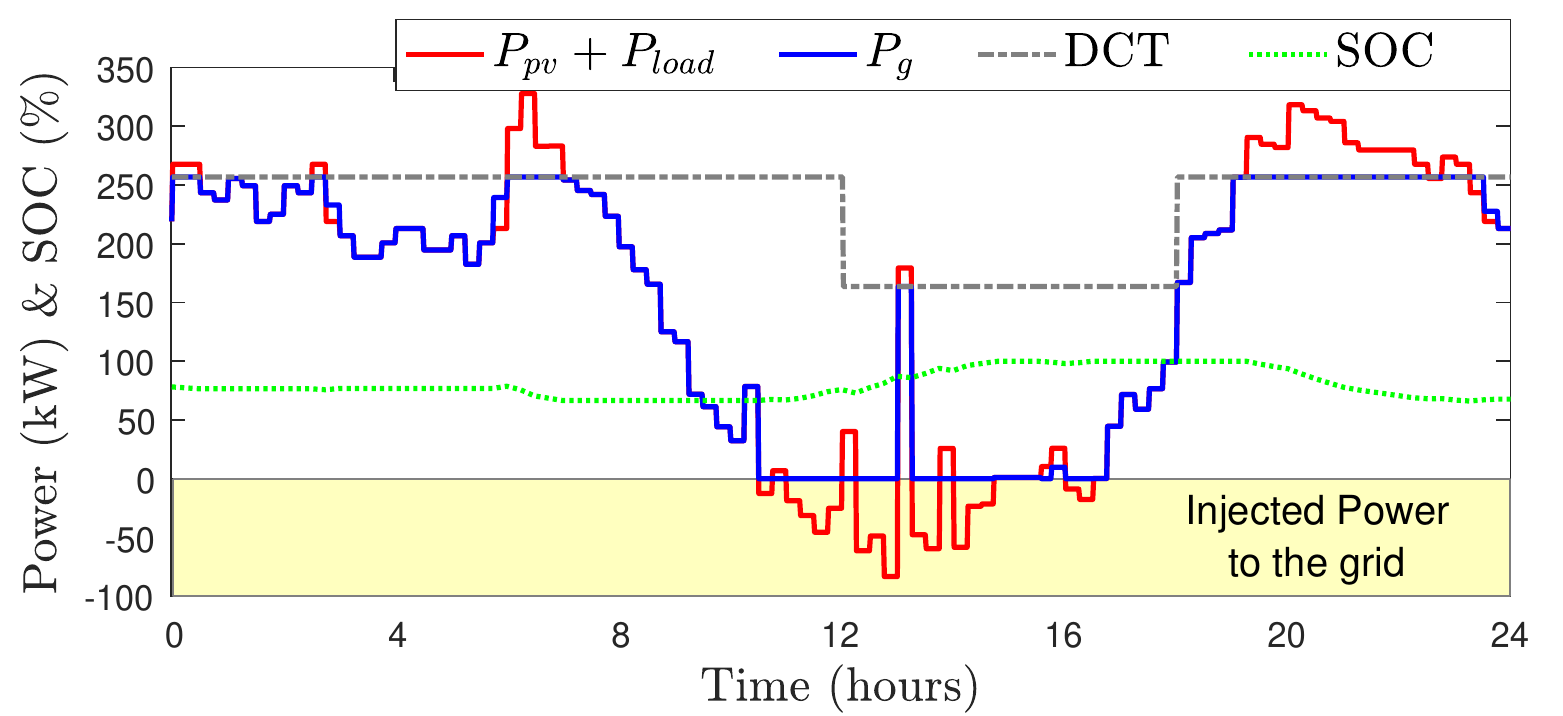}   \vspace{-0.2cm}
	\caption{Sample of a day in July for grocery load while using the proposed MPC-based controller.}  
	\label{fig:Daily_OP}  \vspace{-0.3cm}
\end{figure}
An overview of daily BESS operation and resulting profiles for grocery load (highest variations) with rule-based controller and the proposed MPC-based controller is presented in Figs. \ref{fig:Daily_RB} and \ref{fig:Daily_OP}, respectively. In case of rule-based controller, the battery is fully charged after the peak shaving event in the morning and consequently cannot absorb the excess PV production. However, the proposed method significantly reduces the excess PV injection to the grid by scheduling the BESS to charge during excess generation and discharge during Peak demand. Employing this method significantly improves the annual PV-utilization from $0.008\%$ in case of rule-based controller to $71.15\%$ using the proposed controller (refer to Table \ref{TABLE:Year}). Thus, the effectiveness of the proposed method is evident in reducing the excess PV injection and tracking the DCT. \\ \indent
The monthly effects of reducing excess PV injection are further illustrated in Figs. \ref{fig:Psell_RB} and  \ref{fig:Psell_OP}. It clearly indicates that the excess PV injection to the grid reduces significantly as duration and magnitude of $P_{g}^{sell}$ decreases with MPC-based controller. Table \ref{TABLE:Year} summarizes the annual simulation results for different load profiles. The results reveal that the proposed method was able to significantly enhance the PV-utilization rate between $60$ to $80\%$ while slightly decreasing the demand charge savings to less than $2.5\%$. Moreover, this method also reduces the average annual SOC and has the added benefits of increasing the battery lifetime. It can also be seen that the hospital load has the highest PV-utilization rate because its profile is more aligned with PV demand pattern.
\subsection{Quantitative analysis: optimization time horizon}
To evaluate the effects of optimization horizon, analysis were repeated for different time windows. For instance, we pick the Grocery load profile and the annual results are shown in Table \ref{TABLE:Time}. This table reveals the performance of the proposed method in terms of DC saving (which is the primary application) is not sensitive to optimization horizon; however; increasing the horizon can improve the overall PV-utilization ratio from $59\%$ for 3-hours to $80\%$ for 5-hours time horizon. Generally, choosing a proper time horizon is a trade-off between forecast accuracy and the information needed for planning. Moreover, maintaining $SOC_{req}$ will improve the robustness against PV/load forecast errors by avoiding full discharge of the battery caused by inaccurate estimation.
\subsection{Quantitative analysis: battery size}
Different battery sizes have been considered to evaluate the effects on the performance of the proposed method in case of Grocery load. The results in Table \ref{TABLE:Battery} reveal that increasing the battery size will improve the PV-utilization rate. Generally, this is expected as bigger batteries have higher capacities to shift the load and excess PV-generation.
\begin{table}[!t]
	\centering \vspace{-0.3cm}
	\caption{Analyzing the Effects of Optimization Time Horizon.}
	\label{TABLE:Time}
	\scriptsize
	\renewcommand{\arraystretch}{1.2}
	\begin{tabular}{|c|c|c|c|}
		\hline
		& Time Horizon ($T$)  & DC Saving  ($\%$)     & PV-util. ($\%$)  \\ \hline
		\multirow{3}{*}{\begin{tabular}[c]{@{}c@{}} MPC-based Controller \end{tabular}} 
		&   3-Hours (12 steps)     &     17.12 &  59.55     \\ \cline{2-4} 
		&   4-Hours (16 steps)     &     17.42 &  71.15           \\ \cline{2-4} 
		&   5-Hours (20 steps)     &     17.91 &  80.21         \\ \hline
		\begin{tabular}[c]{@{}c@{}}Rule-based  controller\end{tabular}       &   -    &       18.28    	&   0.008     \\ \hline
	\end{tabular}
\end{table}
\begin{table}[!t]
	\centering \vspace{-0.3cm}
	\caption{Analyzing the Effects of Battery Size.}
	\label{TABLE:Battery}
	\scriptsize
	\renewcommand{\arraystretch}{1.2}
	\begin{tabular}{|c|c|c|c|c|}
		\hline
		\multirow{2}{*}{Battery Size}   & \multicolumn{2}{c|}{Rule-based Controller}    & \multicolumn{2}{c|}{MPC-based Controller}               \\ \cline{2-5} 
		&\!\!\!\!  DC Saving  ($\%$) \!\!\!\! & \!\!\!\!  PV-util. ($\%$)  \!\!\!\!   & \!\!\!\!  DC Saving  ($\%$) \!\!\!\!  & \!\!\!\!  PV-util. ($\%$)  \!\!\!\!    \\ \hline
		\!\!\!\! 280 kW, 170 kWh   \!\!\!\!          &   14.77  &  0.008         &   15.14            &    49.60            \\ \hline
		\!\!\!\! 710 kW, 340 kWh   \!\!\!\!          &   18.28  &   0.008         &   17.42            &  71.15              \\ \hline
		\!\!\!\! 710 kW, 510 kWh   \!\!\!\!          &   19.90  &   0.16          &   17.30            &  75.84              \\ \hline
	\end{tabular}
\end{table}
\section{Conclusions}
In this paper, a new MPC-based controller is proposed to provide benefits in reducing demand charge and increase PV-utilization simultaneously for commercial and industrial customers. This can improve the economic benefits of battery storage systems by stacking multiple services to increase their revenues. The proposed MPC method calculates the optimal charging/discharging profiles for 15 minutes interval which will then be sent to the real-time controller engine. Different load patterns have been used to demonstrate the performance throughout the year. Simulation results demonstrated that the proposed controller provides satisfactory performance by improving the PV-utilization rate between $60$ to $80\%$ while slightly decreasing the demand charge savings (less than $2.5\%$) compared to a demand charge only controller. Quantitative analysis for PV-utilization as a function of battery size and prediction time window has also been carried out.
\bibliography{IEEEabrv_Ehsan,Ref}  
\bibliographystyle{IEEEtran}
\end{document}